\documentclass[11pt]{article}
\usepackage{amsmath,amssymb}
\usepackage{color}
\usepackage{graphicx}
\usepackage{hyperref}
\usepackage{float}
\usepackage{cite}
\usepackage[utf8]{inputenc}
\usepackage{tikz}

\newcommand{\tr}{\mathrm{Tr}\,}

\newcommand{\be}{\begin{equation}}
	\newcommand{\ee}{\end{equation}}
\newcommand{\bea}{\begin{eqnarray}}
	\newcommand{\eea}{\end{eqnarray}}
\newcommand{\ba}{\begin{align}}
	\newcommand{\ea}{\end{align}}
\newcommand{\p}{\partial}

\def \be {\begin{equation}}
	\def \ee {\end{equation}}
\def \ba {\begin{array}}
	\def \ea {\end{array}}
\def \bea{\begin{eqnarray}}
	\def \eea{\end{eqnarray}}

\def \a {\alpha}
\def \b {\beta}

\def \G {\Gamma}
\def \d {\delta}

\def \l {\lambda}

\def \Tr {\mathrm{Tr}\,}

\begin{document}
	
	\thispagestyle{empty}

	\renewcommand{\thefootnote}{\fnsymbol{footnote}}

	\begin{flushright}\footnotesize
		
		\text{CJQS-2024-???} \\
		\text{USTC-ICTS/PCFT-23-27}
		
	\end{flushright}	
	\begin{center}
		{\Large\textbf{\mathversion{bold}
				More Fermionic Supersymmetric Wilson Loops in Four Dimensions
			}
			\par}
		
		\vspace{0.5cm}
		
		\textrm{Hao Ouyang$^1$,~~
			Jun-Bao Wu$^{2, 3}$\footnote{Corresponding author.}}
		\vspace{4mm}
		
		{\small
			\textit{
				$^{1}$Center for Theoretical Physics and College of Physics, Jilin University, Changchun 130012, P.~R.~China,\\		
				$^{2}$Center for Joint Quantum Studies and Department of Physics, School of Science,\\
				Tianjin University, 135 Yaguan Road, Tianjin 300350, P.~R.~China\\
				$^{3}$Peng Huanwu Center for Fundamental Theory, 96 Jinzhai Road, Hefei, Anhui 230026, P. R. China }\\
			\texttt{haoouyang@jlu.edu.cn, junbao.wu@tju.edu.cn}
		}

		

\par\vspace{1.5cm}
		
\textbf{Abstract} \vspace{3mm}
		
\begin{minipage}{\textwidth}
We construct supersymmetric fermionic Wilson loops along general curves in four-dimensional $\mathcal{N}=4$ super Yang-Mills theory and along general planar curves in $\mathcal{N}=2$ superconformal $SU(N)\times SU(N)$ quiver theory. These loops are generalizations of the  Zarembo loops and are cohomologically equivalent to them. In $\mathcal{N}=4$ super Yang-Mills theory, we compute their expectation values and verify the cohomological equivalence relation up to the order $g^4$ in perturbation theory.
\end{minipage}
		
\end{center}
	
\vspace{1.5cm}
	
\newpage
\tableofcontents
\renewcommand{\thefootnote}{\arabic{footnote}}
\setcounter{footnote}{0}


	\section{Introduction}

Bogomol\`nyi-Prasad-Sommerfield (BPS) Wilson loops (WLs)~\cite{hep-th/9803002, hep-th/9803001}  in  four-dimensional $\mathcal{N}=4$ super Yang-Mills theory (SYM) play an important role in the precise checks
of the AdS/CFT correspondence~\cite{hep-th/9711200, hep-th/9802109, hep-th/9802150} since the early days. One of the precise checks is about the vacuum expectation value (vev) of a circular half-BPS  WL in the fundamental representation in $\mathcal{N}=4$ SYM with gauge group $SU(N)$. 	 Based on one-loop computations, it was conjectured~\cite{Erickson:2000af} that, the planar limit of this vev can be obtained from a resummation of the ladder planar diagrams in the Feynman gauge. This conjecture leads to the result that this vev can be computed using a Gaussian matrix model in its own planar limit~\cite{Erickson:2000af}. This vev is a non-trivial function of the 't~Hooft coupling constant $\lambda$ and $N$. The large $N$, large $\lambda$ limit of this vev matches precisely with the prediction from the dual string theory~\cite{hep-th/9809188, hep-th/9904191} using certain half-BPS F-string solutions in the $AdS_5\times S^5$ background. The conjecture about the reduction to the Gaussian matrix model was later proved by Pestun~\cite{0712.2824}  using supersymmetric localization. 	This precise matching between the strong coupling result in the field theory side and the weakly coupled string theory result is among the earliest non-trivial  validations of the AdS/CFT correspondence, extending beyond the checks about correlation functions of BPS local operators related to various non-renormalization theorems~\cite{Lee:1998bxa, Drukker:2009sf, Baggio:2012rr}.
	
Many BPS WLs with fewer supersymmetries were constructed in $\mathcal{N}=4$ SYM. In Zarembo's construction~\cite{Zarembo:2002an}, the loops inside a $\mathbb{R}^n$ subspace of $\mathbb{R}^4$ Euclidean space preserve $1/2^n$ of the Poincar\'{e} supercharges. We will refer to such loops as $1/2^n$ Poincar\'{e} BPS or just $1/2^n$-BPS. By direct perturbative computations, Zarembo found that the leading and next-to-leading corrections to the vev of $1/4$-BPS Zarembo loop vanishes  in the large $N$ limit.  Subsequent arguments were presented to support the result that the vev of any Zarembo loop equals unity exactly even at finite $N$~\cite{Guralnik:2003di, Guralnik:2004yc, Dymarsky:2006ve}. The holographic description of Zarembo loops using calibrated surfaces~\cite{Dymarsky:2006ve} also supports this result.

Another class of BPS WLs was found by Drukker, Giombi, Ricci and Trancanelli (DGRT)~\cite{Drukker:2007dw, Drukker:2007qr}. For an arbitrary curve in $S^3$ they found the suitable scalar coupling to the WL such that the WL preserves at least two linear combinations of Poincar\'e and conformal supercharges. Different from the case of Zarembo loops, the generic DGRT loop has nontrivial vev. It was found~\cite{Drukker:2007yx, Drukker:2007qr} that, when the DGRT loop is restricted to an $S^2$ submanifold, its vev can be obtained from the vev of certain ordinary WL in two-dimensional (non-supersymmetric) Yang-Mills theory on $S^2$ restricted to the zero-instanton sector~\cite{Staudacher:1997kn, Bassetto:1998sr}\footnote{This restriction  leads to the Wu-Mandelstam-Leibbrandt prescription of reguralization~\cite{Wu:1977hi, Mandelstam:1982cb, Leibbrandt:1983pj}.}. A similar relation was also obtained for correlation functions of such DGRT loops and certain local operators on the same $S^2$~\cite{Giombi:2009ds}. Certain classification of BPS WLs in ${\mathcal N}=4$ SYM was performed in~\cite{Dymarsky:2009si}.

The above BPS WLs in ${\mathcal N}=4$ SYM usually involve suitable coupling of scalars to the WLs in the construction. In the study of BPS WLs in three-dimensional super-Chern-Simons theories, fermionic BPS WLs were also constructed. In such construction~\cite{Drukker:2009hy}, the WL couples to the fermions in the theory as well, besides the gauge fields and scalars. The introduction of the fermionic BPS WLs was to resolve a puzzle about the duality between  BPS WLs in ABJM theory~\cite{Aharony:2008ug} and the probe F-string theory in the dual type IIA string theory on $AdS_4\times \mathbb {CP}^3$ background. $1/6$-BPS bosonic WLs in ABJM theory was constructed in~\cite{Drukker:2008jm, Chen:2008bp, Rey:2008bh}. However there are probe F-string solutions~\cite{Drukker:2008jm,  Rey:2008bh} in $AdS_4\times \mathbb {CP}^3$ which are half-BPS and quite reasonably dual to WLs. But no such half-BPS WLs are found among the above $1/6$-BPS bosonic WLs. The construction of half-BPS fermionic WLs by Drukker and Trancanelli~\cite{Drukker:2009hy}   satisfactorily resolved this problem. Later $1/6$-BPS fermionic WLs in ABJM theory were constructed in~\cite{Ouyang:2015iza, Ouyang:2015bmy}. For special choice of the parameters in the constructions, such WLs will go back to the previously found $1/6$-BPS bosonic WLs or half-BPS WLs. It was proposed in \cite{Correa:2019rdk} that a generic 1/6-BPS fermionic WL is dual to an F-string with supersymmetric mixed boundary conditions.

One naturally wonders whether similar BPS fermionic WLs exist in four-dimensional superconformal gauge theories.
In \cite{Ouyang:2022vof} we provided a positive answer to this question by explicitly constructing in four-dimensional $\mathcal{N}=2$ superconformal $SU(N)\times SU(N)$ quiver theory and $\mathcal{N}=4$ SYM. In each theory, we constructed two types of fermionic BPS WLs that preserve some supersymmetries. The first type consists of WLs along an infinite timelike straight line in Lorentzian signature, which preserve one or more Poincar\'{e} supercharges. The second type consists of WLs along a circular contour in Euclidean signature, which preserve one or more linear combinations of Poincar\'{e} and conformal supercharges.
Every fermionic WL belongs to the same $Q$-cohomology class as a bosonic half-BPS WL that shares the same supercharge $Q$. If we assume that this cohomological
equivalence also holds true at the quantum level, we can predict that the fermionic BPS WL has the same expectation value as the bosonic one.

The aim of the present work is to investigate fermionic BPS WLs along more general contours by employing Zarembo's construction. One notable feature of the fermionic Zarembo loops is that the connections are supersymmetric invariant, whereas the supersymmetric variations of the connections of typical fermionic BPS WLs constructed previously are covariant total derivatives\footnote{ The connection in this covariant derivative is just the connection used to define the WL.}. The number of preserved supersymmetries by the WL depends on the choice of the contour and parameters in the connection.
Similar to previously constructed WLs, there exists a cohomological equivalence relation between the fermionic and bosonic Zarembo loops. In $\mathcal{N}=4$ SYM, we verify the cohomological equivalence relation up to order $g^4$ in the perturbation theory.
Our results provide new insights into the structure and properties of BPS WLs in four-dimensional superconformal gauge theories.

The paper is organized as follows.
In section~\ref{section2} we review Zarembo's construction. Then we present our construction of fermionic Zarembo loops in $\mathcal{N}=4$ SYM and compute their expectation values to order $g^4$ at finite $N$.
In section~\ref{N2Z} we construct fermionic Zarembo loops in $\mathcal{N}=2$ superconformal $SU(N)\times SU(N)$ quiver theory and discuss their supersymmetry properties.
Section~\ref{section4} concludes with some remarks.
Appendix~\ref{B} contains some conventions and useful formulas.

	\section{Fermionic supersymetric WLs in $\mathcal{N}=4$ SYM}\label{section2}
	\subsection{Zarembo loop}\label{N4Z}
Let us begin by briefly reviewing Zarembo’s construction \cite{Zarembo:2002an}.
The Euclidean action of $\mathcal{N}=4$ SYM with gauge group $SU(N)$ is
	\begin{equation}
		S_{{\mathcal N}=4} =\int_{{\mathbf R}^4} d^4 x
		\left(\frac{1}{2}{\tr}(F_{MN}F^{MN})
		+i\tr (\Psi^c\Gamma^M D_M\Psi)\right).
	\end{equation}
The $\Gamma^{M}$'s are ten-dimensional gamma matrices.
We use  the index conventions $M,N=0,...,9$ and $I,J=4,...,9$. $\Psi$ satisfies the chiral condition $\Gamma^{0123456789}\Psi=\Psi$ and  $\Psi^c = \Psi^T C$ is the  charge conjugation of $\Psi$. For more detailed conventions, please refer to Appendix \ref{B}.
The action is invariant under the superconformal transformations:
	\begin{equation}
		\begin{split}\label{N4susy}
			& \d A_M=-i \xi^c \G_M\Psi, \\
			& \d \Psi=\frac{1}{2}F_{MN}\Gamma^{MN}\xi-2\Gamma^I A_I \vartheta.	
		\end{split}
	\end{equation}
where $\xi=\theta+x^\mu \Gamma_\mu \vartheta$ with $\mu=0,...,3$. $\xi$ satisfies the chiral condition $\Gamma^{0123456789}\xi=\xi$. The constant spinors $\theta$ and $\vartheta$ generate Poincar\'{e} supersymmetry transformations and  conformal supersymmetry transformations respectively. In $\mathcal{N}=4$ SYM,
a natural generalization of the ordinary WL is the Maldacena-Wilson loop:
	\begin{equation}
		W=\frac{1}{N}\tr \mathcal{P} \exp \left( i g\int_C d \tau   ( A_\mu \dot x^\mu(\tau)+i |\dot x| \Theta^I (\tau) A_I)\right) .
	\end{equation}
Local supersymmetry requires the norm of $\Theta^I$ to be one. One simple example of a globally supersymmetric WL is the one with $C$  a straight line and  $\Theta^{I}$'s being constants. A remarkable generalization has been proposed by Zarembo \cite{Zarembo:2002an}. The Zarembo loops are
defined by
		\begin{equation}
		\Theta^I (\tau)=M^I_\mu \frac{\dot x^\mu}{|\dot x|} .
	    \end{equation}
where $M^I_\mu$ is a constant projection matrix:
	\begin{equation}
		M^I_\mu M^J_\nu\delta_{IJ}=\delta_{\mu \nu}.
	\end{equation}
Considering Poincar\'{e} supersymmetry variation of the loop, some supersymmetries will be preserved if the equation
	\begin{equation}
			 \dot x^\mu \left(\Gamma _\mu+i \Gamma _I M_\mu^I\right)\theta=0,
	\end{equation}
has nontrivial solutions.
When the contour is a generic
curve in $\mathbb{R}^4$,
$\theta$ satisfies four constraints
\begin{equation}
\left(\Gamma _\mu+i \Gamma _I M_\mu^I\right)\theta=0,~~~\mu=0,1,2,3,
\end{equation}
and the WL is $1/16$ Poincar\'{e} BPS.
When the contour of the loop is inside a subspace $\mathbb{R}^n$, there are $n$ independent constraints and the loop is $1/2^n$   BPS\footnote{ Precisely speaking, Zarembo only investigated  the Poincar\'e supercharges preserved by these WLs.  The counting of the supercharges in this and the next subsections is only for Poincar\'{e} supercharges as well.}. Especially, if the operator lies on a straight line there is only one constraint equation on $\theta$ so it is 1/2 BPS.
 Zarembo loop operators on non-straight curves can only be constructed in Euclidean signature. The reason is that if there is more than one constraint equation,  at least one of them corresponds to a space direction and contradicts the real condition of the spinor in Lorentz signature \cite{Ouyang:2015ada}.
Therefore in this work,  we focus on WL operators in Euclidean space.
	
	\subsection{Fermionic loop}\label{subsection2.2}
We now generalize Zarembo’s construction by including coupling to the fermionic fields in ${\mathcal{N}=4}$ SYM. The connection contains both bosonic and fermionic components. BPS fermionic WLs along a straight line were constructed in \cite{Ouyang:2022vof}.
The fermionic component can be obtained by acting on  a specific linear combination of the scalars with a supersymmetry generator $Q_s$ that preserves the loop.
The fermionic supersymmetry generator $Q_s$ is defined as $\delta_{\theta}=  \chi Q_s $, where the bosonic spinor $s$ is related to $\theta $ as $\theta= \chi s$ and $\chi$ is a real Grassmann variable.

To construct a BPS WL on a non-straight curve, we start with a bosonic connection
\begin{equation}
	L_{\mathrm{bos}}=g\dot x^{\mu} (A_\mu+i M_{\mu}^I A_I),
\end{equation}
which is $Q_s$-invariant. When $s$ is constrained by at least two projection equations, i.e.
	\begin{equation}\label{twoconstrains}
		\left(\Gamma _{\mu_1}+i \Gamma _I M _{\mu_1}^I\right)s=\left(\Gamma _{\mu_2}+i \Gamma _I M _{\mu_2}^I\right)s=0,
	\end{equation}
one can prove that 
	\begin{equation}
		s^c \Gamma_M s=0,
	\end{equation}
by using the $SO(4)\times SO(6)$ symmetry (and parity invariance if needed) to transform $M_{\mu}^I$ to a simple form $\delta^{\mu}_{I-4}$.
Therefore we find
	\begin{equation}
	Q_s^2 A_M=Q_s (-i s^c \G_M\Psi) =0.
\end{equation}
A supersymmetric fermionic loop can be constructed as
	\begin{equation}
		W_{\mathrm {fer}}
		=\frac{1}{N}\tr \mathcal{P} \exp \left( i \int_C L d\tau  \right),
	\end{equation}
	with a $Q_s$-invariant connection
		\begin{equation}
L=L_{\mathrm{bos}}+g|\dot{x}|Q_s(m^I(\tau) A_I)=L_{\mathrm{bos}}-ig|\dot{x}|m^I(\tau)s^c \G_{I}\Psi.
	\end{equation}
For the BPS WL constructed in \cite{Ouyang:2022vof}, its connection transforms under supersymmetry as a covariant derivative. However, for the BPS WL we constructed here, its connection itself is supersymmetric invariant. Therefore, we can directly use the trace to construct the WL, without the need to construct supermatrices and take the supertrace as in the case of \cite{Ouyang:2022vof}.

	When the contour is a generic curve in $\mathbb{R}^4$, the WL is 1/16 BPS. When the loop lies in a subspace, we would like to
	find all the $u$ such that $Q_uL=0$.
	To be concrete, we take $M_{\mu}^I=\delta_{\mu}^{I-4}$.
	When the WL is inside the 01 plane,
	 $s$ is constrained by two projection equations:
	\begin{equation}\label{0415}
		(1+i \Gamma_{04})s=(1+i \Gamma_{15})s=0,
	\end{equation}
	and $u$ satisfies the same constraints because $Q_uL_{\mathrm{bos}}=0$.
	When $m^{4}=m^{5}=0$, $Q_uL=Q_uQ_s(m^I A_I)=0$ requires
	\begin{equation}
		m^I s^c \Gamma_{45I} u=0,
	\end{equation}
so the solution space of $u$ is three-dimensional and the WL is 3/16 BPS.	
Otherwise, the solution is $u \propto s$, which can be derived from the vanishing of the $F_{0P}$ and $F_{1P}$ terms, and the WL is 1/16 BPS.
	
When WLs lie along the $012$ subspace,  $s$ and $u$ are constrained by three projection equations.
Therefore $s^c \Gamma_{IJK} u\neq 0$ only if $\{I,J,K\}=\{4,5,6\}$ and $s^c \Gamma_{456} u=0$ only if $u\propto s$.
So when $m^{4}=m^{5}=m^{6}=0$, the WL is 1/8 BPS. Otherwise, the WL is 1/16 BPS.

\subsection{Expectation values in perturbation theory}
The fermionic Zarembo loop is classically  $Q_s$-cohomological equivalent to the bosonic one:
\begin{equation}
	\frac{1}{N}\tr \mathcal{P} \exp \left( i \int_C L d\tau  \right)-\frac{1}{N}\tr \mathcal{P} \exp \left( i \int_C L_{\mathrm {bos}} d\tau \right)=Q_s V,
\end{equation}	
where $V$ can be constructed as
\begin{equation}
	\begin{split}
		V&=\frac{1}{N}\sum_{n=1}^\infty 	\mathrm{Tr}\mathcal{P}\Bigg(e^{i  \int L_{\mathrm{bos}} d\tau} \int_{\tau_1>\tau_2>...>\tau_n}d\tau_1 d\tau_2...d\tau_n \Lambda(\tau_1)F(\tau_2)...F(\tau_n)\Bigg),\\
		\Lambda&=m^I A_I,~~~ F=Q_s \Lambda.
	\end{split}
\end{equation}	
If 	the $Q_s$-cohomological equivalence holds at the quantum level, the expectation values of the bosonic and fermionic loops should be equal.	
The	expectation values of the bosonic Zarembo loops are known to be  exactly one~\cite{Guralnik:2003di,Guralnik:2004yc,Dymarsky:2006ve}.
In this subsection, we compute the expectation value of the fermionic Zarembo loop to order $g^4$ in perturbation theory to test the $Q_s$-cohomological equivalence.  We will use  regularization by dimensional reduction~\cite{Siegel:1979wq} as in~\cite{Erickson:2000af}. We do not take the planar limit in this computation.

Let us first review the calculation of the vacuum expectation value of  the bosonic loop in~ \cite{Zarembo:2002an}.
In the Feynman gauge, the tree-level propagators take the form
\begin{align}
\label{Dx}\langle A(x)_M^a A(y)_N^b\rangle_0
&=\delta^{ab}\delta_{MN}D(x-y),\\
\langle \Psi(x)^a \bar\Psi(y)^b\rangle_0
&=i\delta^{ab} \Gamma^\mu \partial_\mu D(x-y).
\end{align}
Although the explicit forms of the propagators will not be necessary for our discussion below, we give the  tree level and one-loop corrected propagators in  regularization by dimensional reduction for completeness in Appendix \ref{B} and the explicit form of $D(x)$ in $2\omega$ dimensions is
\begin{equation}
  D(x)=  \frac{ \Gamma(\omega-1)}{4\pi^\omega}\frac{1}{ (x^2)^{\omega-1} } .
\end{equation}

\begin{figure}
\begin{center}

\begin{tikzpicture}[x=0.6pt,y=0.6pt,yscale=-1,xscale=1]

\draw  [color={rgb, 255:red, 74; green, 144; blue, 226 }  ,draw opacity=1 ] (270,88.5) .. controls (270,56.74) and (295.74,31) .. (327.5,31) .. controls (359.26,31) and (385,56.74) .. (385,88.5) .. controls (385,120.26) and (359.26,146) .. (327.5,146) .. controls (295.74,146) and (270,120.26) .. (270,88.5) -- cycle ;
\draw    (327,32) .. controls (328.67,33.67) and (328.67,35.33) .. (327,37) .. controls (325.33,38.67) and (325.33,40.33) .. (327,42) .. controls (328.67,43.67) and (328.67,45.33) .. (327,47) .. controls (325.33,48.67) and (325.33,50.33) .. (327,52) .. controls (328.67,53.67) and (328.67,55.33) .. (327,57) .. controls (325.33,58.67) and (325.33,60.33) .. (327,62) .. controls (328.67,63.67) and (328.67,65.33) .. (327,67) .. controls (325.33,68.67) and (325.33,70.33) .. (327,72) .. controls (328.67,73.67) and (328.67,75.33) .. (327,77) .. controls (325.33,78.67) and (325.33,80.33) .. (327,82) .. controls (328.67,83.67) and (328.67,85.33) .. (327,87) .. controls (325.33,88.67) and (325.33,90.33) .. (327,92) .. controls (328.67,93.67) and (328.67,95.33) .. (327,97) .. controls (325.33,98.67) and (325.33,100.33) .. (327,102) .. controls (328.67,103.67) and (328.67,105.33) .. (327,107) .. controls (325.33,108.67) and (325.33,110.33) .. (327,112) .. controls (328.67,113.67) and (328.67,115.33) .. (327,117) .. controls (325.33,118.67) and (325.33,120.33) .. (327,122) .. controls (328.67,123.67) and (328.67,125.33) .. (327,127) .. controls (325.33,128.67) and (325.33,130.33) .. (327,132) .. controls (328.67,133.67) and (328.67,135.33) .. (327,137) .. controls (325.33,138.67) and (325.33,140.33) .. (327,142) -- (327,146) -- (327,146) ;

\draw (312.4,5.4) node [anchor=north west][inner sep=0.75pt]    {$L_{\mathrm{bos}}$};
\draw (313.4,151.4) node [anchor=north west][inner sep=0.75pt]    {$L_{\mathrm{bos}}$};

\end{tikzpicture}

\end{center}
\caption{Feynman diagram for the bosonic loop at order $g^2$.}
\label{fig:1loopbos}	
\end{figure}
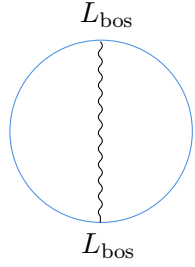

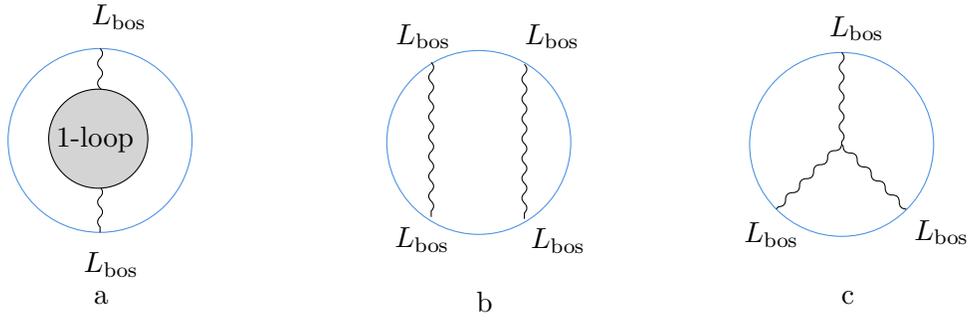
\begin{figure}
\begin{center}

\begin{tikzpicture}[x=0.75pt,y=0.75pt,yscale=-1,xscale=1]

\draw  [color={rgb, 255:red, 74; green, 144; blue, 226 }  ,draw opacity=1 ] (56,73.4) .. controls (56,47.77) and (76.77,27) .. (102.4,27) .. controls (128.03,27) and (148.8,47.77) .. (148.8,73.4) .. controls (148.8,99.03) and (128.03,119.8) .. (102.4,119.8) .. controls (76.77,119.8) and (56,99.03) .. (56,73.4) -- cycle ;
\draw    (102.4,27) .. controls (104.07,28.67) and (104.07,30.33) .. (102.4,32) .. controls (100.73,33.67) and (100.73,35.33) .. (102.4,37) .. controls (104.07,38.67) and (104.07,40.33) .. (102.4,42) .. controls (100.73,43.67) and (100.73,45.33) .. (102.4,47) .. controls (104.07,48.67) and (104.07,50.33) .. (102.4,52) .. controls (100.73,53.67) and (100.73,55.33) .. (102.4,57) .. controls (104.07,58.67) and (104.07,60.33) .. (102.4,62) .. controls (100.73,63.67) and (100.73,65.33) .. (102.4,67) .. controls (104.07,68.67) and (104.07,70.33) .. (102.4,72) .. controls (100.73,73.67) and (100.73,75.33) .. (102.4,77) .. controls (104.07,78.67) and (104.07,80.33) .. (102.4,82) .. controls (100.73,83.67) and (100.73,85.33) .. (102.4,87) .. controls (104.07,88.67) and (104.07,90.33) .. (102.4,92) .. controls (100.73,93.67) and (100.73,95.33) .. (102.4,97) .. controls (104.07,98.67) and (104.07,100.33) .. (102.4,102) .. controls (100.73,103.67) and (100.73,105.33) .. (102.4,107) .. controls (104.07,108.67) and (104.07,110.33) .. (102.4,112) .. controls (100.73,113.67) and (100.73,115.33) .. (102.4,117) -- (102.4,119.8) -- (102.4,119.8) ;
\draw  [draw opacity=0][fill={rgb, 255:red, 212; green, 212; blue, 212 }  ,fill opacity=1 ] (76.5,72.5) .. controls (76.5,58.69) and (87.69,47.5) .. (101.5,47.5) .. controls (115.31,47.5) and (126.5,58.69) .. (126.5,72.5) .. controls (126.5,86.31) and (115.31,97.5) .. (101.5,97.5) .. controls (87.69,97.5) and (76.5,86.31) .. (76.5,72.5) -- cycle ;

\draw  [color={rgb, 255:red, 74; green, 144; blue, 226 }  ,draw opacity=1 ] (247,74.4) .. controls (247,48.77) and (267.77,28) .. (293.4,28) .. controls (319.03,28) and (339.8,48.77) .. (339.8,74.4) .. controls (339.8,100.03) and (319.03,120.8) .. (293.4,120.8) .. controls (267.77,120.8) and (247,100.03) .. (247,74.4) -- cycle ;
\draw    (269.4,34) .. controls (271.07,35.67) and (271.07,37.33) .. (269.4,39) .. controls (267.73,40.67) and (267.73,42.33) .. (269.4,44) .. controls (271.07,45.67) and (271.07,47.33) .. (269.4,49) .. controls (267.73,50.67) and (267.73,52.33) .. (269.4,54) .. controls (271.07,55.67) and (271.07,57.33) .. (269.4,59) .. controls (267.73,60.67) and (267.73,62.33) .. (269.4,64) .. controls (271.07,65.67) and (271.07,67.33) .. (269.4,69) .. controls (267.73,70.67) and (267.73,72.33) .. (269.4,74) .. controls (271.07,75.67) and (271.07,77.33) .. (269.4,79) .. controls (267.73,80.67) and (267.73,82.33) .. (269.4,84) .. controls (271.07,85.67) and (271.07,87.33) .. (269.4,89) .. controls (267.73,90.67) and (267.73,92.33) .. (269.4,94) .. controls (271.07,95.67) and (271.07,97.33) .. (269.4,99) .. controls (267.73,100.67) and (267.73,102.33) .. (269.4,104) .. controls (271.07,105.67) and (271.07,107.33) .. (269.4,109) -- (269.4,112) -- (269.4,112) ;
\draw    (316.4,35) .. controls (318.07,36.67) and (318.07,38.33) .. (316.4,40) .. controls (314.73,41.67) and (314.73,43.33) .. (316.4,45) .. controls (318.07,46.67) and (318.07,48.33) .. (316.4,50) .. controls (314.73,51.67) and (314.73,53.33) .. (316.4,55) .. controls (318.07,56.67) and (318.07,58.33) .. (316.4,60) .. controls (314.73,61.67) and (314.73,63.33) .. (316.4,65) .. controls (318.07,66.67) and (318.07,68.33) .. (316.4,70) .. controls (314.73,71.67) and (314.73,73.33) .. (316.4,75) .. controls (318.07,76.67) and (318.07,78.33) .. (316.4,80) .. controls (314.73,81.67) and (314.73,83.33) .. (316.4,85) .. controls (318.07,86.67) and (318.07,88.33) .. (316.4,90) .. controls (314.73,91.67) and (314.73,93.33) .. (316.4,95) .. controls (318.07,96.67) and (318.07,98.33) .. (316.4,100) .. controls (314.73,101.67) and (314.73,103.33) .. (316.4,105) .. controls (318.07,106.67) and (318.07,108.33) .. (316.4,110) -- (316.4,113) -- (316.4,113) ;
\draw  [color={rgb, 255:red, 74; green, 144; blue, 226 }  ,draw opacity=1 ] (430,75.4) .. controls (430,49.77) and (450.77,29) .. (476.4,29) .. controls (502.03,29) and (522.8,49.77) .. (522.8,75.4) .. controls (522.8,101.03) and (502.03,121.8) .. (476.4,121.8) .. controls (450.77,121.8) and (430,101.03) .. (430,75.4) -- cycle ;
\draw    (476.4,29) .. controls (478.07,30.67) and (478.07,32.33) .. (476.4,34) .. controls (474.73,35.67) and (474.73,37.33) .. (476.4,39) .. controls (478.07,40.67) and (478.07,42.33) .. (476.4,44) .. controls (474.73,45.67) and (474.73,47.33) .. (476.4,49) .. controls (478.07,50.67) and (478.07,52.33) .. (476.4,54) .. controls (474.73,55.67) and (474.73,57.33) .. (476.4,59) .. controls (478.07,60.67) and (478.07,62.33) .. (476.4,64) .. controls (474.73,65.67) and (474.73,67.33) .. (476.4,69) .. controls (478.07,70.67) and (478.07,72.33) .. (476.4,74) -- (476.4,75) -- (476.4,75) ;
\draw    (476.4,75) .. controls (476.4,77.36) and (475.22,78.54) .. (472.86,78.54) .. controls (470.51,78.54) and (469.33,79.72) .. (469.33,82.07) .. controls (469.33,84.43) and (468.15,85.61) .. (465.79,85.61) .. controls (463.44,85.61) and (462.26,86.79) .. (462.26,89.14) .. controls (462.26,91.5) and (461.08,92.68) .. (458.72,92.68) .. controls (456.37,92.68) and (455.19,93.86) .. (455.19,96.21) .. controls (455.19,98.57) and (454.01,99.75) .. (451.65,99.75) .. controls (449.3,99.75) and (448.12,100.93) .. (448.12,103.28) .. controls (448.12,105.64) and (446.94,106.82) .. (444.58,106.82) -- (443.4,108) -- (443.4,108) ;
\draw    (509,108) .. controls (506.64,108) and (505.46,106.82) .. (505.46,104.46) .. controls (505.46,102.11) and (504.28,100.93) .. (501.93,100.93) .. controls (499.57,100.93) and (498.39,99.75) .. (498.39,97.39) .. controls (498.39,95.04) and (497.21,93.86) .. (494.86,93.86) .. controls (492.5,93.86) and (491.32,92.68) .. (491.32,90.32) .. controls (491.32,87.97) and (490.14,86.79) .. (487.79,86.79) .. controls (485.43,86.79) and (484.25,85.61) .. (484.25,83.25) .. controls (484.25,80.9) and (483.07,79.72) .. (480.72,79.72) .. controls (478.36,79.72) and (477.18,78.54) .. (477.18,76.18) -- (476.4,75.4) -- (476.4,75.4) ;

\draw (93,128.4) node [anchor=north west][inner sep=0.75pt]    {$L_{\mathrm{bos}}$};
\draw (97,3.4) node [anchor=north west][inner sep=0.75pt]    {$L_{\mathrm{bos}}$};
\draw (79,65) node [anchor=north west][inner sep=0.75pt]   [align=left] {1-loop};
\draw (249.8,116.2) node [anchor=north west][inner sep=0.75pt]    {$L_{\mathrm{bos}}$};
\draw (318.4,116.4) node [anchor=north west][inner sep=0.75pt]    {$L_{\mathrm{bos}}$};
\draw (250.4,13.4) node [anchor=north west][inner sep=0.75pt]    {$L_{\mathrm{bos}}$};
\draw (315.4,13.4) node [anchor=north west][inner sep=0.75pt]    {$L_{\mathrm{bos}}$};
\draw (426,113.4) node [anchor=north west][inner sep=0.75pt]    {$L_{\mathrm{bos}}$};
\draw (469,9.4) node [anchor=north west][inner sep=0.75pt]    {$L_{\mathrm{bos}}$};
\draw (511.8,112.2) node [anchor=north west][inner sep=0.75pt]    {$L_{\mathrm{bos}}$};
\draw (98,148.42) node [anchor=north west][inner sep=0.75pt]   [align=left] {a};
\draw (291,148.42) node [anchor=north west][inner sep=0.75pt]   [align=left] {b};
\draw (475,148.42) node [anchor=north west][inner sep=0.75pt]   [align=left] {c};

\end{tikzpicture}

\end{center}
\caption{Feynman diagrams for the bosonic loop at order $g^4$. Diagram (b) represents all distinct types of contractions including non-planar contributions.}
\label{fig:2loopbos}	
\end{figure}

At order $g^2$, the Feynman diagram depicted in figure \ref{fig:1loopbos} vanishes because
\begin{equation}
\tr\langle
L_{\mathrm{bos}}(x_1)L_{\mathrm{bos}}(x_2)  \rangle_0
\propto
\dot x_1^\mu \dot x_2^\nu(\delta_{\mu\nu}-\delta_{IJ}M^{I}_\mu M^{J}_\nu)D(x_1-x_2)=0.
\end{equation}
Because of the same reason, diagram (b) in figure {\ref{fig:2loopbos}} is zero.
The one-loop propagators are
\begin{align}
\langle A(x)_\mu^a A(y)_\nu^b\rangle_1
&=g^2 N \delta^{ab}  \frac{\Gamma(\omega-1)\Gamma(\omega-2)}
	{32\pi^{2\omega}(2\omega-3)}
\left(\frac{ \delta_{\mu\nu}}{((x-y)^2)^{2\omega-3}}
-\frac{\partial_\mu \partial_\nu (((x-y)^2)^{4-2 \omega })}{8 (\omega -3) (\omega -2)}
\right),\\
\langle A(x)_I^a A(y)_J^b\rangle_1
&=g^2 N \delta^{ab}  \delta_{IJ} \frac{\Gamma(\omega-1)\Gamma(\omega-2)}
	{32\pi^{2\omega}(2\omega-3)}
	\frac{1}{((x-y)^2)^{2\omega-3}},\\
\langle \Psi(x)^a \bar\Psi(y)^b\rangle_1
&=	-i g^2 N \delta^{ab}  \frac{\Gamma(\omega-1)\Gamma(\omega-2)}
	{8\pi^{2\omega}}
	\frac{(x^\mu-y^\mu) \Gamma_\mu}{((x-y)^2)^{2\omega-2}}.
\end{align}
Because the one-loop scalar and vector propagators are equal up to a total derivative, diagram (a) in figure {\ref{fig:2loopbos}} vanishes.
To compute diagram (c), we use
\begin{equation}
\langle
\tr(L_{\mathrm{bos}}(x_1)L_{\mathrm{bos}}(x_2) L_{\mathrm{bos}}(x_3) )
\tr(\partial_M A_N(x) [A^M(x), A^N(x)])\rangle_0
=0,
\end{equation}
where the convention $\partial_I=0$ is used. Here when one $L_{\mathrm{bos}}$ is contracted with $\partial_M A_N$ and another $L_{\mathrm{bos}}$  with $A^N$, one can find that the result is proportional to 
$\delta_{\mu}^N\delta_{\nu N}-M^{I}_\mu M^{J}_\nu\delta_{IJ}=0$ and thus  diagram (c) vanishes.
Therefore the vev of the bosonic loop equals unity up to order $g^4$.

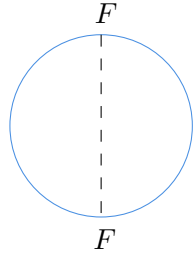
\begin{figure}
\begin{center}
	\begin{tikzpicture}[x=0.6pt,y=0.6pt,yscale=-1,xscale=1]
	
	\draw  [color={rgb, 255:red, 74; green, 144; blue, 226 }  ,draw opacity=1 ] (270,88.5) .. controls (270,56.74) and (295.74,31) .. (327.5,31) .. controls (359.26,31) and (385,56.74) .. (385,88.5) .. controls (385,120.26) and (359.26,146) .. (327.5,146) .. controls (295.74,146) and (270,120.26) .. (270,88.5) -- cycle ;
	\draw  [dash pattern={on 4.5pt off 4.5pt}]  (327.5,31) -- (327.5,146) ;
	
	\draw (322,9.4) node [anchor=north west][inner sep=0.75pt]    {$F$};
	\draw (321,152.4) node [anchor=north west][inner sep=0.75pt]    {$F$};

\end{tikzpicture}
\end{center}
\caption{Feynman diagram involving $F$ at order $g^2$.}
\label{fig:1loopfer}	
\end{figure}

For the fermionic loop, we need to consider diagrams with fermion insertions.	We assume that the parameters $m^I$'s are independent of $g$.
At order $g^2$, the Feynman diagram depicted in figure \ref{fig:1loopfer} vanishes:
	\begin{equation}
		\begin{split}
			&\tr\langle \int d\tau_{1>2}
			g m^I(\tau_1)\bar{s}\Gamma _I\Psi(\tau_1)\,
			g m^J(\tau_2)\bar{\Psi }(\tau_2)\Gamma _J s        \rangle_0\\
			=&ig^2N \int d\tau_{1>2}
			m^I(\tau_1)m^J(\tau_2)\bar{s}\Gamma _I\Gamma^\mu \Gamma _J s \partial_\mu D(x_1-x_2)\\
			=&0,
		\end{split}
	\end{equation}
	where we have used $s^c \Gamma_M s=0$ and $s^c  \Gamma_{M_1M_2M_3} s=0$.

\begin{figure}

\begin{tikzpicture}[x=0.75pt,y=0.75pt,yscale=-1,xscale=1]
			
			\draw  [color={rgb, 255:red, 74; green, 144; blue, 226 }  ,draw opacity=1 ] (52,80.4) .. controls (52,54.77) and (72.77,34) .. (98.4,34) .. controls (124.03,34) and (144.8,54.77) .. (144.8,80.4) .. controls (144.8,106.03) and (124.03,126.8) .. (98.4,126.8) .. controls (72.77,126.8) and (52,106.03) .. (52,80.4) -- cycle ;
			\draw  [dash pattern={on 4.5pt off 4.5pt}]  (98.4,34) -- (98.4,126.8) ;
			\draw  [draw opacity=0][fill={rgb, 255:red, 212; green, 212; blue, 212 }  ,fill opacity=1 ] (72.5,79.5) .. controls (72.5,65.69) and (83.69,54.5) .. (97.5,54.5) .. controls (111.31,54.5) and (122.5,65.69) .. (122.5,79.5) .. controls (122.5,93.31) and (111.31,104.5) .. (97.5,104.5) .. controls (83.69,104.5) and (72.5,93.31) .. (72.5,79.5) -- cycle ;
			
			\draw  [color={rgb, 255:red, 74; green, 144; blue, 226 }  ,draw opacity=1 ] (487,82.4) .. controls (487,56.77) and (507.77,36) .. (533.4,36) .. controls (559.03,36) and (579.8,56.77) .. (579.8,82.4) .. controls (579.8,108.03) and (559.03,128.8) .. (533.4,128.8) .. controls (507.77,128.8) and (487,108.03) .. (487,82.4) -- cycle ;
			\draw    (533.4,36) .. controls (535.07,37.67) and (535.07,39.33) .. (533.4,41) .. controls (531.73,42.67) and (531.73,44.33) .. (533.4,46) .. controls (535.07,47.67) and (535.07,49.33) .. (533.4,51) .. controls (531.73,52.67) and (531.73,54.33) .. (533.4,56) .. controls (535.07,57.67) and (535.07,59.33) .. (533.4,61) .. controls (531.73,62.67) and (531.73,64.33) .. (533.4,66) .. controls (535.07,67.67) and (535.07,69.33) .. (533.4,71) .. controls (531.73,72.67) and (531.73,74.33) .. (533.4,76) .. controls (535.07,77.67) and (535.07,79.33) .. (533.4,81) -- (533.4,82) -- (533.4,82) ;
			\draw  [dash pattern={on 4.5pt off 4.5pt}]  (533.4,82) -- (518.6,96.8) -- (500.4,115) ;
			\draw  [dash pattern={on 4.5pt off 4.5pt}]  (566,115) -- (533.4,82.4) ;
			\draw  [color={rgb, 255:red, 74; green, 144; blue, 226 }  ,draw opacity=1 ] (198,82.4) .. controls (198,56.77) and (218.77,36) .. (244.4,36) .. controls (270.03,36) and (290.8,56.77) .. (290.8,82.4) .. controls (290.8,108.03) and (270.03,128.8) .. (244.4,128.8) .. controls (218.77,128.8) and (198,108.03) .. (198,82.4) -- cycle ;
			\draw  [dash pattern={on 4.5pt off 4.5pt}]  (220.4,42) -- (220.4,120) ;
			\draw    (267.4,43) .. controls (269.07,44.67) and (269.07,46.33) .. (267.4,48) .. controls (265.73,49.67) and (265.73,51.33) .. (267.4,53) .. controls (269.07,54.67) and (269.07,56.33) .. (267.4,58) .. controls (265.73,59.67) and (265.73,61.33) .. (267.4,63) .. controls (269.07,64.67) and (269.07,66.33) .. (267.4,68) .. controls (265.73,69.67) and (265.73,71.33) .. (267.4,73) .. controls (269.07,74.67) and (269.07,76.33) .. (267.4,78) .. controls (265.73,79.67) and (265.73,81.33) .. (267.4,83) .. controls (269.07,84.67) and (269.07,86.33) .. (267.4,88) .. controls (265.73,89.67) and (265.73,91.33) .. (267.4,93) .. controls (269.07,94.67) and (269.07,96.33) .. (267.4,98) .. controls (265.73,99.67) and (265.73,101.33) .. (267.4,103) .. controls (269.07,104.67) and (269.07,106.33) .. (267.4,108) .. controls (265.73,109.67) and (265.73,111.33) .. (267.4,113) .. controls (269.07,114.67) and (269.07,116.33) .. (267.4,118) -- (267.4,121) -- (267.4,121) ;
			\draw  [color={rgb, 255:red, 74; green, 144; blue, 226 }  ,draw opacity=1 ] (339,81.73) .. controls (339,56.11) and (359.77,35.33) .. (385.4,35.33) .. controls (411.03,35.33) and (431.8,56.11) .. (431.8,81.73) .. controls (431.8,107.36) and (411.03,128.13) .. (385.4,128.13) .. controls (359.77,128.13) and (339,107.36) .. (339,81.73) -- cycle ;
			\draw  [dash pattern={on 4.5pt off 4.5pt}]  (361.4,41.33) -- (361.4,119.33) ;
			\draw  [dash pattern={on 4.5pt off 4.5pt}]  (408.4,42.33) -- (408.4,120.33) ;
			
			\draw (94,128.4) node [anchor=north west][inner sep=0.75pt]    {$F$};
			\draw (93,18.4) node [anchor=north west][inner sep=0.75pt]    {$F$};
			\draw (483,120.4) node [anchor=north west][inner sep=0.75pt]    {$F$};
			\draw (526,16.4) node [anchor=north west][inner sep=0.75pt]    {$L_{\mathrm{bos}}$};
			\draw (568.8,119.2) node [anchor=north west][inner sep=0.75pt]    {$F$};
			\draw (200.8,124.2) node [anchor=north west][inner sep=0.75pt]    {$F$};
			\draw (269.4,124.4) node [anchor=north west][inner sep=0.75pt]    {$L_{\mathrm{bos}}$};
			\draw (201.4,21.4) node [anchor=north west][inner sep=0.75pt]    {$F$};
			\draw (266.4,21.4) node [anchor=north west][inner sep=0.75pt]    {$L_{\mathrm{bos}}$};
			\draw (75,72) node [anchor=north west][inner sep=0.75pt]   [align=left] {1-loop};
			\draw (341.8,123.53) node [anchor=north west][inner sep=0.75pt]    {$F$};
			\draw (410.4,123.73) node [anchor=north west][inner sep=0.75pt]    {$F$};
			\draw (343.4,20.73) node [anchor=north west][inner sep=0.75pt]    {$F$};
			\draw (407.4,20.73) node [anchor=north west][inner sep=0.75pt]    {$F$};
			\draw (94,146.67) node [anchor=north west][inner sep=0.75pt]   [align=left] {a};
			\draw (243,146.67) node [anchor=north west][inner sep=0.75pt]   [align=left] {b};
			\draw (384,146.67) node [anchor=north west][inner sep=0.75pt]   [align=left] {c};
			\draw (535,146.67) node [anchor=north west][inner sep=0.75pt]   [align=left] {d};

\end{tikzpicture}
		\caption{Feynman diagrams involving $F$ at order $g^4$. Diagrams (b) and (c) represent all distinct types of contractions including non-planar contributions.}
		\label{fig:2loopfer}
	\end{figure}
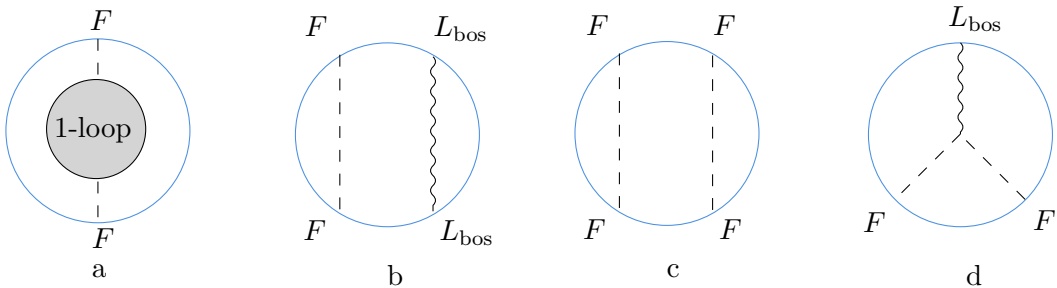

The order $g^4$  Feynman diagrams involving $F=-im^I\bar{s}\Gamma _I\Psi$ are shown in figure \ref{fig:2loopfer}.
Since the one-loop fermion propagator is proportional to $x^\mu \Gamma_\mu$ as the tree-level one, diagram (a) does not contribute.
Diagrams (b) and (c) vanish for the same reason. Diagram (d) vanishes because it contains the following structure:
	\begin{equation}
	\dot{x}^\mu	s^c \Gamma^{K} \Gamma_\rho (\Gamma_\mu+i M_{\mu}^I \Gamma_I) \Gamma_\nu \Gamma^{J} s =0,
	\end{equation}
where we have used $s^c \Gamma_M s=s^c \Gamma_{M_1M_2M_3} s=0$ and   the anti-communication relations of the gamma matrices to move $ (\Gamma_\mu+i M_{\mu}^I \Gamma_I) $ to the place just before $s$. Then   $ \dot{x}^\mu (\Gamma_\mu+i M_{\mu}^I \Gamma_I) s=0$ has been used. Therefore the expectation value of the fermionic Zarembo loop is trivial and the $Q_s$-cohomological equivalence is confirmed up to order $g^4$.

\subsection{WL with conformal supersymmetries}
The WL constructed above typically does not preserve conformal supersymmetry. However, for a circular contour, the WL may preserve conformal symmetry. In this section, we will provide an example which is a generalization of one special $1/4$-BPS bosonic loop studied in \cite{Drukker:2006ga} being also a special case of Zarembo loops.
Let us consider WLs on a circle  $(x^0, x^1,x^2,x^3)=r(\cos \tau,\sin \tau,0,0)$.
We start with the bosonic connection:
	\begin{equation}
		L_{\mathrm{bos}}=g (\dot x^\mu A_\mu+i r \cos \tau A_4+i r \sin \tau A_5).	
	\end{equation}
Referring to the Poincar\'{e} supersymmetry generator notation, we employ $S_v$ with a bosonic spinor $v$ to represent the superconformal generator. The  preserved supersymmetries by $L_{\mathrm{bos}}$ satisfy
	\begin{equation}
		(-\sin \tau\Gamma_0+\cos \tau\Gamma_1+i  \cos \tau \Gamma_4+i  \sin \tau \Gamma_5)(s+r(\cos\tau\Gamma_0+\sin \tau\Gamma_1)v)=0.	
	\end{equation}
The result is
	\begin{equation}
		(1+i \Gamma_{14})s= (1-i \Gamma_{05})s=(1+i \Gamma_{14})v= (1-i \Gamma_{05})v=0,	
	\end{equation}
so $L_{\mathrm{bos}}$  preserves four Poincar\'{e} supersymmetries and four conformal supersymmetries.
	We take $F=g|\dot{x}|m^6(\tau) Q_s A_6=-i g r m^6(\tau) s^c \Gamma_6 \Psi $. As discussed in subsection \ref{subsection2.2}, the connection $L=L_{\mathrm{bos}}+F$ preserves $Q_u$ with
$s^c \Gamma_{456}u=0$.
	
	Acting $S_v$ on $F$ we get
	\begin{equation}
		\begin{split}
			S_v F=&-\frac{i}{2}g r m^6 s^c\Gamma_{6}\Gamma^{MN}   (r\cos\tau\Gamma_0+r\sin \tau\Gamma_1)v F_{MN}+2igrm^6 s^c\Gamma_{6}\Gamma^{J}v A_J.
		\end{split}
	\end{equation}
We find $F$  preserves $S_v$ with $v \propto \Gamma_{6}s$. Therefore the WL associated with the connection $L$ preserves three Poincar\'{e} supercharges and one conformal supercharge, so it is $1/8$-BPS.

	\section{Fermionic Zarembo loops in $\mathcal{N}=2$ superconformal $SU(N)\times SU(N)$ quiver theory}\label{N2Z}
In this section, we construct fermionic Zarembo loops in the $\mathcal{N}=2$ superconformal $SU(N)\times SU(N)$ quiver theory which can be obtained via orbifolding $\mathcal{N}=4$ SYM by $\mathbb{Z}_2$.
There are two $\mathcal{N}=2$ vector multiplets for the two gauge group factors.
The component fields can be arranged into $2\times 2$ block matrices:
\begin{equation}
	\begin{split}
		&A_m=\left( \ba{cc} A_m^{(1)} &0\\  0&  A_m^{(2)}\ea\right),~~~
		m=0,...,5,\\
		&\l_\alpha=\left( \ba{cc} \l_\alpha^{(1)} &0\\ 0&  \l_\alpha^{(2)}\ea\right),~~~
		\alpha=1,2,
	\end{split}
\end{equation}
where $A_\mu$ with $\mu=0,...,3$ is the gauge field and $A_{4,5}$ are two real scalars.
We use six-dimensional spinorial notations for the spinors.
The gaugino fermions $\l_1$ and $\l_2$ are $SO(6)$ Weyl spinors of chirality $-1$ for $\Gamma^{012345}$.
There are also two bifundamental hypermultiplets with  component fields:
\begin{equation}
	q^\a=\left( \ba{cc} 0 & q^{(1)\a}\\ q^{(2)\a}&0\ea\right),~~~\psi=\left( \ba{cc} 0 & \psi^{(1)}\\ \psi^{(2)}&0\ea\right),	
\end{equation}
where $q^{1,2}$ are complex scalars and   $\psi$ is an $SO(6)$ Weyl spinor of chirality $+1$ for $\Gamma^{012345}$.
The Euclidean action of the ${\cal N}=2$ gauge theory is
\bea
S_{{\mathcal N}=2} &= &\int d^4 x
\left(\tr(\frac{1}{2}F_{mn}F^{mn}
{+i \bar\lambda^\alpha \Gamma^m D_m\lambda_\alpha}
{+2D_m q_\alpha D^m q^\alpha}
{+2i \bar\psi \Gamma^m D_m \psi}\right.\nonumber\\
&& {-2\sqrt{2}g \bar\lambda^{\alpha A} q_\alpha T_a\psi}
{+2\sqrt{2}g \bar\psi T_a q^\alpha \lambda^A_\alpha)}
{+2g^2 \tr(q_\alpha T^a q^\beta)
	\tr(q_\beta T_a q^\alpha)}\nonumber\\ && \left.{- g^2 \tr(q_\alpha T_a q^\alpha)
	\tr(q_\beta T^a q^\beta)}\right)\, ,\eea
where $T^a$ are the generators of the gauge group.
We define $\bar \lambda^\alpha$ as
$\bar \lambda^\alpha=-\epsilon^{\alpha\beta}\l^c_\beta$ where $\epsilon^{\alpha\beta}$ is the antisymmetric symbol with $\epsilon^{12}=1$.
 The fermions $\psi$ and $\bar \psi$ are independent.
The $\mathcal{N}=2$ superconformal symmetry is still preserved when one leaves the orbifold point by independently varying the coupling constants for the two factors of the gauge group. The two coupling constants can be written as:
\begin{equation}
	g=\left( \ba{cc} g^{(1)} I_N &0\\  0&  g^{(2)}I_N\ea\right),
\end{equation}
where $I_N$ denotes the $N \times N$ identity matrix. The definitions of the covariant derivatives are
\bea
D_\mu \lambda&=&\p_\mu \lambda-i g [A_\mu, \lambda] ,\\
D_\mu q^\alpha&=&\partial_\mu q^\alpha-i { g} A_\mu q^\alpha,\\
D_\mu q_\alpha&=&\partial_\mu q_\alpha+i { g}q_\alpha A_\mu,\\
D_\mu \Psi&=&\partial_\mu \Psi-i {g} A_\mu \Psi.
\eea
The  $\mathcal{N}=2$  superconformal transformations are:
\begin{equation}
	\begin{split}
		& \d A_m=-i \bar\xi^\a\G_m\l_\a=i \bar\lambda^\a\G_m\xi_\a,  \\
		& \d q^\alpha=-i\sqrt{2}\bar\xi^\a\psi, \\
		& \d q_\a=-i\sqrt{2}\bar\psi \xi_\a , \\
		& \d\l_\alpha^A={\frac{1}{2}F^A_{m n} \G^{mn}\xi_\alpha}
		{+2ig   q_\a T^A q^\b  \xi_\b
			-ig   q_\b T^A q^\b \xi_\a}
		{-2 A^A_a \Gamma^a \vartheta_\alpha}, \\
		& \d\bar\l^{\alpha A}= {-\frac{1}{2}\bar \xi^\alpha F^A_{m n} \G^{mn}}
		{-2ig   q_\b T^A q^\a  \bar\xi^\b
			+ig   q_\b T^A q^\b \bar\xi^\a}
		{+2 \bar\vartheta^\alpha A^A_a \Gamma^a} , \\
		& \d\psi=-\sqrt{2}D_m q^\a \G^m \xi_\a {-2\sqrt{2} q^\a\vartheta_\alpha},\\
		& \d\bar\psi=\sqrt{2}\bar \xi ^\a \G^m D_m q_\a {-2\sqrt{2} \bar\vartheta^\alpha q_\a},
	\end{split}
\end{equation}
where $\xi_\alpha=\theta_\alpha+x^\mu \Gamma_\mu \vartheta_\alpha$ has chirality $-1$ for $\Gamma^{012345}$ and the index $a=4,5$. The constant spinors $\theta_\alpha$ and $\vartheta_\alpha$ are parameters associated with Poincar\'{e} supersymmetry and conformal supersymmetry respectively.

	There are only two real adjoint scalars in the $\mathcal{N}=2$ theory. One can define	
	a planar bosonic Zarembo loop:
	\begin{equation}
		W_{\mathrm {bos}}=\frac{1}{2N}\tr\mathcal{P} e^{i\int \mathrm{d}\tau L_{\mathrm{bos}}(\tau)},~~~
		L_{\mathrm{bos}}=g\dot x^\mu (A_\mu+i M_{\mu}^a A_a).
	\end{equation}
Without loss of generality, we choose a contour inside the 01 plane and take $M_{\mu}^a=\delta^a_{\mu+4}$.
	The Poincar\'{e} supersymmetries preserved by this bosonic WL satisfy
	\begin{equation}
		\dot x^\mu \left(\Gamma _\mu+i \Gamma _a M_\mu^a\right)\theta_\alpha=0.
	\end{equation}
	Parameterizing $\theta_\a$ as $\theta_\a= \chi s_\a$ with a real Grassmann number $\chi$, we find
	\begin{equation}\label{saequation}
	\Big(\frac{1}{2}+\frac{i}{2} \Gamma _{04}\Big)s_\alpha=\Big(\frac{1}{2}+\frac{i}{2} \Gamma _{15}\Big)s_\alpha=0.
	\end{equation}
For each $\alpha$ there is only one linearly independent solution and $s_1 \propto s_2$. So the WL preserves two supersymmetries.

The connection of the fermionic loop can be constructed as a supermatrix
\begin{equation}
	L=L_{\mathrm{bos}}+F,
\end{equation}
where the fermionic part takes the form:
\begin{align}
	F= g|\dot{x}|(\zeta^c  \psi+ \bar\psi  \eta), ~~~
	\zeta^c = \left( \ba{cc} \zeta^{(1)c}  I_N &0 \\  0& \zeta^{(2)c} I_N \ea\right), ~~~
	\eta=\left( \ba{cc} \eta^{(2)}I_N &0 \\  0& \eta^{(1)}I_N \ea\right).
\end{align}	
The bosonic spinors $\zeta$ and $\eta$ satisfy
\begin{equation}
	\Big(\frac{1}{2}+\frac{i}{2} \Gamma _{04}\Big)\zeta
	=\Big(\frac{1}{2}+\frac{i}{2} \Gamma _{15}\Big)\zeta
	=\Big(\frac{1}{2}+\frac{i}{2} \Gamma _{04}\Big)\eta
	=\Big(\frac{1}{2}+\frac{i}{2} \Gamma _{15}\Big)\eta=0,
\end{equation}
and they can depend on $\tau$.

We define the fermionic supersymmetry generator $Q_s$ as $\delta_{\theta}= \sqrt{2}\chi Q_s $ by extracting a real Grassmann number $\chi$.
For any $s_\alpha$ satisfying (\ref{saequation}), we have $Q_s F=0$.
Therefore we can define a BPS WL by using a trace or supertrace:
\begin{equation}
W_{\mathrm{tr}}=	\frac{1}{2N}\tr \mathcal{P} \exp \left( i \int_C L d\tau  \right),~~~\mathrm{or}~~
W_{\mathrm{str}}=	\frac{1}{2N}\mathrm{sTr} \mathcal{P} \exp \left( i \int_C L d\tau  \right).
\end{equation}
Both of them preserve two Poincar\'{e} supersymmetries.
It is straightforward to show  that $F$ can be constructed by acting $Q_s$ on a linear combination of $q_\alpha$ and $q^\alpha$, so the
The fermionic Zarembo loops are classically  $Q_s$-cohomological equivalent to the bosonic one.

\section{Conclusion and discussions}\label{section4}

In this paper, we have constructed fermionic Zarembo loops in four-dimensional $\mathcal{N}=4$ SYM and $\mathcal{N}=2$ superconformal $SU(N)\times SU(N)$ quiver theories. These loops are generalizations of the bosonic Zarembo loops. In the construction, we strongly made use of special properties of Poincar\'{e} supercharges preserved by the bosonic Zarembo loops.
We examined how the choice of contour and connection parameters affects the number of supersymmetries preserved by the fermionic Zarembo loops. We have shown that the fermionic Zarembo loops are cohomologically equivalent to the bosonic ones and computed their expectation values in perturbation theory up to order $g^4$ in $\mathcal{N}=4$ SYM. We have also discussed the possibility of preserving conformal supercharges.

Our results provide new examples of BPS WLs in four-dimensional superconformal gauge theories.
They also raise some open questions and potential extensions of our work.
It would be interesting to study the holographic duals of the fermionic Zarembo loops in IIB superstring theory on $AdS_5 \times S^5$ or its orbifold background.
It would be worthwhile to consider other generalizations of known fermionic BPS WLs such as possible fermionic  DGRT loops.

Both $\mathcal{N}=4$ SYM and ABJM theories are integrable in the planar limit. If we insert a composite operator inside the WL, the WL provides boundary conditions/interactions for the open spin chain from the composite operator. Half-BPS WLs in both theories lead to integrable open spin chains~\cite{Drukker:2006xg, Correa:2018fgz, Correa:2023lsm}. The correlation function of a WL and a single trace operator which is an eigenstate of the dilatation operator is proportional to the overlap of a boundary state and a Bethe state. For half-BPS WLs in the antisymmetric representation in ${\cal N}=4$ theory, such boundary states are integrable in the planar limit~\cite{JKV}. For bosonic $1/6$-BPS WLs and half-BPS WLs in the fundamental representation in  ABJM theory, such boundary states are integrable at least at tree level in the planar limit~\cite{Jiang:2023cdm}. It is interesting to study whether the fermionic BPS WLs constructed in~\cite{Ouyang:2022vof}
and this paper can lead to integrable open chains and/or integrable boundary
states.

\section*{Acknowledgments}
We would like to thank Ziwen Kong and Zhengbin Yang for very helpful discussions. 	The work of HO is supported by the National Natural Science Foundation of China, Grant No.~12205115. The work of JW  is supported by the National Natural Science Foundation of China, Grant No.~12375066, 11935009, 11975164, 12247103, 12047502, and  Natural Science Foundation of Tianjin under Grant No.~20JCYBJC00910.

\begin{appendix}
\section{Conventions and useful formulas}

\label{B}
The action of $\mathcal{N}=4$ SYM in Euclidean signature is:
\begin{equation}
	S_{\mathcal{N}=4} =\int d^4 x
	\left(\frac{1}{2}{\tr}(F_{MN}F^{MN})
	+i\tr (\bar\Psi\Gamma^M D_M\Psi)\right).
\end{equation}
We split the ten-dimensional indices $M,N,P=0,...9$ into two groups:  $\mu,\nu=0,1,2,3$ and $I,J=4,...,9$.
We denote
$A_M=(A^a_\mu,\Phi_I^a) T^a$, $\Psi=\Psi^a T^a$ and
$\Gamma^M=(\Gamma^\mu,\Gamma^i)$.	In Euclidean signature $\bar\Psi=\Psi^T C$ where $C$ is the charge conjugation matrix.
The field strength is defined as
\be
F_{MN}=\partial_M A_N-\partial_N A_M-i g[A_M, A_N]\,.
\ee
And the covariant derivative is defined as 
\be D_M\Psi=\partial_M\Psi-ig[A_M, \Psi]\,. \ee

Following the conventions in \cite{Ouyang:2022vof}, we use the representation for the ten-dimensional gamma matrices:
\begin{equation}
	\begin{split}
		\Gamma_{(10)}^m&=I_4 \otimes \Gamma_{(6)}^m, ~~~m=0,...,5,\\
		\Gamma_{(10)}^p&=\Gamma_{(4)}^{10-p}\otimes \Gamma_{(6)}^{012345}, ~~~p=6,...,9.
	\end{split}
\end{equation}
The four-dimensional gamma matrices are
\begin{equation}
	\Gamma_{(4)}^{j}=\left(
	\begin{array}{cc}
		0	&  -i\sigma_j\\
		i \sigma_j	& 0
	\end{array}
	\right) ,~~~
	\Gamma_{(4)}^{4}=\left(
	\begin{array}{cc}
		0	&  I_2\\
		I_2	& 0
	\end{array}
	\right),
\end{equation}
 where $\sigma_j$'s denote the Pauli matrices. The six-dimensional gamma matrices are
\begin{equation}
	\begin{split}
	\G^0_{(6)}=&-\sigma_2\otimes  \sigma_3\otimes \sigma_3,\\
	\G^1_{(6)}=&\sigma_1\otimes \sigma_3\otimes \sigma_3,\\
	\G^2_{(6)}=&-I_2 \otimes \sigma_1\otimes  \sigma_3,\\
	\G^3_{(6)}=&-I_2 \otimes \sigma_2\otimes  \sigma_3,\\
	\G^4_{(6)}=&I_2 \otimes I_2 \otimes \sigma_1,\\
	\G^5_{(6)}=&I_2 \otimes I_2\otimes \sigma_2,\\	
	\end{split}
\end{equation}
which were also used in section~\ref{N2Z}.
The charge conjugate matrices can be defined as
\begin{align}
 C_{(4)}&=
 \left(
	\begin{array}{cc}
		i\sigma_2	&  0\\
		0	& i\sigma_2		
	\end{array}
	\right),\\
C_{(6)}&=
\left(
\begin{array}{cccccccc}
 0 & 0 & 0 & 0 & 0 & 0 & 0 & 1 \\
 0 & 0 & 0 & 0 & 0 & 0 & -1 & 0 \\
 0 & 0 & 0 & 0 & 0 & 1 & 0 & 0 \\
 0 & 0 & 0 & 0 & -1 & 0 & 0 & 0 \\
 0 & 0 & 0 & -1 & 0 & 0 & 0 & 0 \\
 0 & 0 & 1 & 0 & 0 & 0 & 0 & 0 \\
 0 & -1 & 0 & 0 & 0 & 0 & 0 & 0 \\
 1 & 0 & 0 & 0 & 0 & 0 & 0 & 0 \\
\end{array}
\right),\\
 C_{(10)}&=C_{(4)}\otimes C_{(6)}.
\end{align}

 The commutation relations and normalization of the generators of the Lie algebra $su(N)$ are,
\begin{equation}
	[ T^a,T^b]=if^{abc}T^c,~~~
	\Tr T^aT^b=\frac{1}{2}\delta^{ab}.
\end{equation}
The  structure constants satisfy the identity
\begin{equation}
	\sum_{c,d}f^{acd}f^{bcd}=N\delta^{ab}.
\end{equation}

Adding terms involving the ghosts and gauge fixing terms to the Euclidean ${\mathcal N}=4$ SYM action, we get
\begin{equation}
	\begin{split}
		S_{\mathcal{N}=4}^{\rm{total}}
		&=\int d^4x \frac{1}{2}\Big( \frac{1}{2}\left(F_{\mu\nu}^a\right)^2
		+\left(\partial_\mu \Phi_i^a+g f^{abc}A^b_\mu \Phi_i^c\right)^2
		+i\bar\Psi^a \Gamma^\mu\left(
		\partial_\mu \Psi^a+g f^{abc}A_\mu^b\Psi^c\right)
		\\&\qquad
		+ig f^{abc}\bar\Psi^a\Gamma^i\Phi^b_i\Psi^c -g^2\sum_{i<j}f^{abc}f^{ade}\Phi_i^b\Phi_j^c\Phi_i^d\Phi_j^e
		+\partial^\mu\bar c^a\left(\partial_\mu c^a+g f^{abc}A_\mu^b c^c\right)
		\\&\qquad
		+\xi( \partial^\mu A_\mu^a)^2\Big).
	\end{split}
	\label{actionn}
\end{equation}

The unrenormalized gluon and scalar propagators up to one-loop order in the Feynman gauge using regularization by dimensional reduction can be found in \cite{Erickson:2000af}.
Up to one loop order,  the gluon propagator in $d=2\omega$ in momentum space is
\begin{equation}
	\Delta^{ab}_{\mu\nu}(p)=\delta^{ab} \frac{\delta_{\mu\nu}}{p^2}-g^2 N
	\frac{\Gamma(2-\omega) \Gamma(\omega)\Gamma(\omega-1)}{(4\pi)^{\omega}
		\Gamma(2\omega)}\cdot 4(2\omega-1)\delta^{ab}\frac{
		\delta_{\mu\nu}-p_\mu p_\nu/p^2}{(p^2)^{3-\omega}}.
\end{equation}
and  the scalar propagator is
\begin{equation}
	D^{ab}_{IJ}(p)=\delta^{ab}
	\frac{\delta_{IJ}}{p^2}-g^2N
	\frac{\Gamma(2-\omega)
		\Gamma(\omega)\Gamma(\omega-1)}{(4\pi)^{\omega} \Gamma(2\omega)}\cdot
	4(2\omega-1)\frac{ \delta_{ij}\delta^{ab}}{(p^2)^{3-\omega}}.
\end{equation}
The  fermion propagator to one loop order can be computed as
\begin{equation}
	\begin{split}
		S^{ab}(p)	&=	-\delta^{ab}\frac{p_\mu \Gamma^\mu}{p^2}-g^2\int \frac{ d^{2 \omega }l}{(2 \pi )^{2 \omega }}
		\frac{-p_\mu \Gamma^\mu}{p^2}
		\left(f^{acd} \Gamma ^M\right)
		\frac{-(p_\nu-l_\nu)  \Gamma^\nu}{(p-l)^2}
		\left(f^{dcb} \Gamma _M\right)
		\frac{ 1}{l^2}
		\frac{-p_\rho \Gamma^\rho}{p^2}   \\
		&=-\delta^{ab}\frac{p_\mu \Gamma^\mu}{p^2}-N\delta^{ab} g^2\int \frac{ d^{2 \omega }l}{(2 \pi )^{2 \omega }}
		\frac{p_\mu \Gamma^\mu}{p^2}
		\Gamma ^M
		\frac{(p_\nu-l_\nu)  \Gamma^\nu}{(p-l)^2}
		\Gamma _M
		\frac{ 1}{l^2}
		\frac{p_\rho \Gamma^\rho}{p^2}   \\
		&=-\delta^{ab}\frac{p_\mu \Gamma^\mu}{p^2}+8 N\delta^{ab} g^2\int \frac{ d^{2 \omega }l}{(2 \pi )^{2 \omega }}
		\frac{p_\mu \Gamma^\mu}{p^2}
		\frac{(p_\nu-l_\nu)  \Gamma^\nu}{(p-l)^2l^2}
		\frac{p_\rho \Gamma^\rho}{p^2}  \\
		&=-\delta^{ab}\frac{p_\mu \Gamma^\mu}{p^2}
		+g^2N\delta^{ab}
		\frac{\Gamma(2-\omega)
			\Gamma(\omega)\Gamma(\omega-1)}{(4\pi)^{\omega} \Gamma(2\omega)}\cdot
		8(2\omega-1)\frac{p_\mu \Gamma^\mu}{(p^2)^{3-\omega}},
	\end{split}
\end{equation}
where we have used
\begin{equation}
	\int \frac{ d^{2 \omega }l}{(2 \pi )^{2 \omega }}
	\frac{1}{\left(l^2+2 l\cdot p+M^2\right)^A}=\frac{\Gamma (A-\omega )}{(4 \pi )^{\omega } \Gamma (A)\left(M^2-p^2\right)^{A-\omega }},
\end{equation}
and the Feynman parameterization formula
\begin{equation}
	\prod_i A_i^{-n_i}=\frac{ \Gamma(\sum n_i)}{\prod_i\Gamma(n_i)}
	\int_0^1 dx_1\cdots dx_k\,x_1^{n_1-1}\cdots x_k^{n_k-1}
	\frac{ \delta(1-\sum x_i) }{\big(\sum_i A_i x_i\big)^{\sum n_i}}.
\end{equation}
The propagators in position space can be obtained via the Fourier transform		
\begin{equation}
	\int \frac{d^{2\omega}p}{(2\pi)^{2\omega} } \frac{ e^{ip\cdot x} }{ p^{2s} }
	=\frac{\Gamma(\omega-s)}{4^s\pi^\omega\Gamma(s) }\frac{1}{(x^2)^{\omega-s}}.
\end{equation}		
The results are
\begin{align}
	\Delta^{ab}_{\mu\nu}(x)&=\delta^{ab} \delta_{\mu\nu}
	\frac{ \Gamma(\omega-1)}{4\pi^\omega}\frac{1}{ (x^2)^{\omega-1} } \nonumber\\&
	+g^2 N \delta^{ab}   \frac{\Gamma(\omega-1)\Gamma(\omega-3)}
	{64\pi^{2\omega}(2\omega-3)}
	\frac{\delta_{ \mu \nu}  x^2 (2 \omega -5)+x_\mu x_\nu (6-4 \omega )}{(x^2)^{2\omega-2}},\label{Dxy}\\
	D^{ab}_{IJ}(x)&=\delta^{ab} \delta_{IJ}
	\frac{ \Gamma(\omega-1)}{4\pi^\omega}\frac{1}{ (x^2)^{\omega-1} }
	+g^2 N \delta^{ab}  \delta_{IJ}\frac{\Gamma(\omega-1)\Gamma(\omega-2)}
	{32\pi^{2\omega}(2\omega-3)}
	\frac{1}{(x^2)^{2\omega-3}},\\
	S^{ab}(x)&=-i\delta^{ab}
	\frac{ \Gamma(\omega)}{2\pi^\omega}\frac{x^\mu \Gamma_\mu}{ (x^2)^{\omega} }
	-i g^2 N \delta^{ab}  \frac{\Gamma(\omega-1)\Gamma(\omega-2)}
	{8\pi^{2\omega}}
	\frac{x^\mu \Gamma_\mu}{(x^2)^{2\omega-2}}.
\end{align}

	\end{appendix}


\begin{thebibliography}{10}

\bibitem{hep-th/9803002}
J.~M. Maldacena, \emph{{Wilson loops in large N field theories}},
  \href{http://dx.doi.org/10.1103/PhysRevLett.80.4859}{\emph{Phys. Rev. Lett.}
  {\bfseries 80} (1998) 4859--4862},
  [\href{https://arxiv.org/abs/hep-th/9803002}{{\ttfamily hep-th/9803002}}].

\bibitem{hep-th/9803001}
S.-J. Rey and J.-T. Yee, \emph{{Macroscopic strings as heavy quarks in large N
  gauge theory and anti-de Sitter supergravity}},
  \href{http://dx.doi.org/10.1007/s100520100799}{\emph{Eur. Phys. J. C}
  {\bfseries 22} (2001) 379--394},
  [\href{https://arxiv.org/abs/hep-th/9803001}{{\ttfamily hep-th/9803001}}].

\bibitem{hep-th/9711200}
J.~M. Maldacena, \emph{{The Large N limit of superconformal field theories and
  supergravity}}, \href{http://dx.doi.org/10.1023/A:1026654312961}{\emph{Adv.
  Theor. Math. Phys.} {\bfseries 2} (1998) 231--252},
  [\href{https://arxiv.org/abs/hep-th/9711200}{{\ttfamily hep-th/9711200}}].

\bibitem{hep-th/9802109}
S.~S. Gubser, I.~R. Klebanov and A.~M. Polyakov, \emph{{Gauge theory
  correlators from noncritical string theory}},
  \href{http://dx.doi.org/10.1016/S0370-2693(98)00377-3}{\emph{Phys. Lett. B}
  {\bfseries 428} (1998) 105--114},
  [\href{https://arxiv.org/abs/hep-th/9802109}{{\ttfamily hep-th/9802109}}].

\bibitem{hep-th/9802150}
E.~Witten, \emph{{Anti-de Sitter space and holography}},
  \href{http://dx.doi.org/10.4310/ATMP.1998.v2.n2.a2}{\emph{Adv. Theor. Math.
  Phys.} {\bfseries 2} (1998) 253--291},
  [\href{https://arxiv.org/abs/hep-th/9802150}{{\ttfamily hep-th/9802150}}].

\bibitem{Erickson:2000af}
J.~K. Erickson, G.~W. Semenoff and K.~Zarembo, \emph{{Wilson loops in N=4
  supersymmetric Yang-Mills theory}},
  \href{http://dx.doi.org/10.1016/S0550-3213(00)00300-X}{\emph{Nucl. Phys. B}
  {\bfseries 582} (2000) 155--175},
  [\href{https://arxiv.org/abs/hep-th/0003055}{{\ttfamily hep-th/0003055}}].

\bibitem{hep-th/9809188}
D.~E. Berenstein, R.~Corrado, W.~Fischler and J.~M. Maldacena, \emph{{The
  operator product expansion for Wilson loops and surfaces in the large N
  limit}}, \href{http://dx.doi.org/10.1103/PhysRevD.59.105023}{\emph{Phys. Rev.
  D} {\bfseries 59} (1999) 105023},
  [\href{https://arxiv.org/abs/hep-th/9809188}{{\ttfamily hep-th/9809188}}].

\bibitem{hep-th/9904191}
N.~Drukker, D.~J. Gross and H.~Ooguri, \emph{{Wilson loops and minimal
  surfaces}}, \href{http://dx.doi.org/10.1103/PhysRevD.60.125006}{\emph{Phys.
  Rev. D} {\bfseries 60} (1999) 125006},
  [\href{https://arxiv.org/abs/hep-th/9904191}{{\ttfamily hep-th/9904191}}].

\bibitem{0712.2824}
V.~Pestun, \emph{{Localization of gauge theory on a four-sphere and
  supersymmetric Wilson loops}},
  \href{http://dx.doi.org/10.1007/s00220-012-1485-0}{\emph{Commun. Math. Phys.}
  {\bfseries 313} (2012) 71--129},
  [\href{https://arxiv.org/abs/0712.2824}{{\ttfamily 0712.2824}}].

\bibitem{Lee:1998bxa}
S.~Lee, S.~Minwalla, M.~Rangamani and N.~Seiberg, \emph{{Three point functions
  of chiral operators in D = 4, N=4 SYM at large N}},
  \href{http://dx.doi.org/10.4310/ATMP.1998.v2.n4.a1}{\emph{Adv. Theor. Math.
  Phys.} {\bfseries 2} (1998) 697--718},
  [\href{https://arxiv.org/abs/hep-th/9806074}{{\ttfamily hep-th/9806074}}].

\bibitem{Drukker:2009sf}
N.~Drukker and J.~Plefka, \emph{{Superprotected n-point correlation functions
  of local operators in N=4 super Yang-Mills}},
  \href{http://dx.doi.org/10.1088/1126-6708/2009/04/052}{\emph{JHEP} {\bfseries
  04} (2009) 052}, [\href{https://arxiv.org/abs/0901.3653}{{\ttfamily
  0901.3653}}].

\bibitem{Baggio:2012rr}
M.~Baggio, J.~de~Boer and K.~Papadodimas, \emph{{A non-renormalization theorem
  for chiral primary 3-point functions}},
  \href{http://dx.doi.org/10.1007/JHEP07(2012)137}{\emph{JHEP} {\bfseries 07}
  (2012) 137}, [\href{https://arxiv.org/abs/1203.1036}{{\ttfamily 1203.1036}}].

\bibitem{Zarembo:2002an}
K.~Zarembo, \emph{{Supersymmetric Wilson loops}},
  \href{http://dx.doi.org/10.1016/S0550-3213(02)00693-4}{\emph{Nucl. Phys. B}
  {\bfseries 643} (2002) 157--171},
  [\href{https://arxiv.org/abs/hep-th/0205160}{{\ttfamily hep-th/0205160}}].

\bibitem{Guralnik:2003di}
Z.~Guralnik and B.~Kulik, \emph{{Properties of chiral Wilson loops}},
  \href{http://dx.doi.org/10.1088/1126-6708/2004/01/065}{\emph{JHEP} {\bfseries
  01} (2004) 065}, [\href{https://arxiv.org/abs/hep-th/0309118}{{\ttfamily
  hep-th/0309118}}].

\bibitem{Guralnik:2004yc}
Z.~Guralnik, S.~Kovacs and B.~Kulik, \emph{{Less is more: Non-renormalization
  theorems from lower dimensional superspace}},
  \href{http://dx.doi.org/10.1142/S0217751X05028193}{\emph{Int. J. Mod. Phys.
  A} {\bfseries 20} (2005) 4546--4553},
  [\href{https://arxiv.org/abs/hep-th/0409091}{{\ttfamily hep-th/0409091}}].

\bibitem{Dymarsky:2006ve}
A.~Dymarsky, S.~S. Gubser, Z.~Guralnik and J.~M. Maldacena, \emph{{Calibrated
  surfaces and supersymmetric Wilson loops}},
  \href{http://dx.doi.org/10.1088/1126-6708/2006/09/057}{\emph{JHEP} {\bfseries
  09} (2006) 057}, [\href{https://arxiv.org/abs/hep-th/0604058}{{\ttfamily
  hep-th/0604058}}].

\bibitem{Drukker:2007dw}
N.~Drukker, S.~Giombi, R.~Ricci and D.~Trancanelli, \emph{{More supersymmetric
  Wilson loops}},
  \href{http://dx.doi.org/10.1103/PhysRevD.76.107703}{\emph{Phys. Rev. D}
  {\bfseries 76} (2007) 107703},
  [\href{https://arxiv.org/abs/0704.2237}{{\ttfamily 0704.2237}}].

\bibitem{Drukker:2007qr}
N.~Drukker, S.~Giombi, R.~Ricci and D.~Trancanelli, \emph{{Supersymmetric
  Wilson loops on $S^3$}},
  \href{http://dx.doi.org/10.1088/1126-6708/2008/05/017}{\emph{JHEP} {\bfseries
  05} (2008) 017}, [\href{https://arxiv.org/abs/0711.3226}{{\ttfamily
  0711.3226}}].

\bibitem{Drukker:2007yx}
N.~Drukker, S.~Giombi, R.~Ricci and D.~Trancanelli, \emph{{Wilson loops: From
  four-dimensional SYM to two-dimensional YM}},
  \href{http://dx.doi.org/10.1103/PhysRevD.77.047901}{\emph{Phys. Rev. D}
  {\bfseries 77} (2008) 047901},
  [\href{https://arxiv.org/abs/0707.2699}{{\ttfamily 0707.2699}}].

\bibitem{Staudacher:1997kn}
M.~Staudacher and W.~Krauth,
\emph{{Two-dimensional QCD in the Wu-Mandelstam-Leibbrandt prescription}},
\href{https://doi.org/10.1103/PhysRevD.57.2456}
{\emph{Phys. Rev. D} \bfseries{57} (1998), 2456-2459}
[\href{https://arxiv.org/abs/hep-th/9709101}{{\ttfamily hep-th/9709101}}].


\bibitem{Bassetto:1998sr}
A.~Bassetto and L.~Griguolo,
\emph{Two-dimensional QCD, instanton contributions and the perturbative Wu-Mandelstam-Leibbrandt prescription},
\href{https://doi.org/10.1016/S0370-2693(98)01319-7}{\emph Phys. Lett. B} {\bfseries{443}} (1998) 325--330
[\href{https://arxiv.org/abs/hep-th/9806037}{{\ttfamily hep-th/9806037}}].


\bibitem{Wu:1977hi}
T.~T. Wu, \emph{{Two-Dimensional Yang-Mills Theory in the Leading 1/N
  Expansion}},
  \href{http://dx.doi.org/10.1016/0370-2693(77)90762-6}{\emph{Phys. Lett. B}
  {\bfseries 71} (1977) 142--144}.

\bibitem{Mandelstam:1982cb}
S.~Mandelstam, \emph{{Light Cone Superspace and the Ultraviolet Finiteness of
  the N=4 Model}},
  \href{http://dx.doi.org/10.1016/0550-3213(83)90179-7}{\emph{Nucl. Phys. B}
  {\bfseries 213} (1983) 149--168}.

\bibitem{Leibbrandt:1983pj}
G.~Leibbrandt, \emph{{The Light Cone Gauge in Yang-Mills Theory}},
  \href{http://dx.doi.org/10.1103/PhysRevD.29.1699}{\emph{Phys. Rev. D}
  {\bfseries 29} (1984) 1699}.

\bibitem{Giombi:2009ds}
S.~Giombi and V.~Pestun, \emph{{Correlators of local operators and 1/8 BPS
  Wilson loops on $S^2$ from 2d YM and matrix models}},
  \href{http://dx.doi.org/10.1007/JHEP10(2010)033}{\emph{JHEP} {\bfseries 10}
  (2010) 033}, [\href{https://arxiv.org/abs/0906.1572}{{\ttfamily 0906.1572}}].

\bibitem{Dymarsky:2009si}
A.~Dymarsky and V.~Pestun, \emph{{Supersymmetric Wilson loops in ${\mathcal
  N}=4$ SYM and pure spinors}},
  \href{http://dx.doi.org/10.1007/JHEP04(2010)115}{\emph{JHEP} {\bfseries 04}
  (2010) 115}, [\href{https://arxiv.org/abs/0911.1841}{{\ttfamily 0911.1841}}].

\bibitem{Drukker:2009hy}
N.~Drukker and D.~Trancanelli, \emph{{A Supermatrix model for $\mathcal{N}=6$ super
  Chern-Simons-matter theory}},
  \href{http://dx.doi.org/10.1007/JHEP02(2010)058}{\emph{JHEP} {\bfseries 02}
  (2010) 058}, [\href{https://arxiv.org/abs/0912.3006}{{\ttfamily 0912.3006}}].

\bibitem{Aharony:2008ug}
O.~Aharony, O.~Bergman, D.~L. Jafferis and J.~Maldacena, \emph{{$\mathcal{N}=6$
  superconformal Chern-Simons-matter theories, M2-branes and their gravity
  duals}}, \href{http://dx.doi.org/10.1088/1126-6708/2008/10/091}{\emph{JHEP}
  {\bfseries 10} (2008) 091},
  [\href{https://arxiv.org/abs/0806.1218}{{\ttfamily 0806.1218}}].

\bibitem{Drukker:2008jm}
N.~Drukker, J.~Plefka and D.~Young,
\emph{{Wilson loops in 3-dimensional $\mathcal{N}=6$ supersymmetric Chern-Simons Theory and their string theory duals}},
\href{https://doi.org/10.1088/1126-6708/2008/11/019}{\emph{JHEP} {\bfseries 11} (2008), 019}
[\href{https://arxiv.org/abs/0809.2787}{{\ttfamily 0809.2787}}].

\bibitem{Chen:2008bp}
B.~Chen and J.-B. Wu, \emph{{Supersymmetric Wilson loops in ${\mathcal N}=6$
  super Chern-Simons-matter theory}},
  \href{http://dx.doi.org/10.1016/j.nuclphysb.2009.09.015}{\emph{Nucl. Phys. B}
  {\bfseries 825} (2010) 38--51},
  [\href{https://arxiv.org/abs/0809.2863}{{\ttfamily 0809.2863}}].

\bibitem{Rey:2008bh}
S.-J. Rey, T.~Suyama and S.~Yamaguchi, \emph{{Wilson Loops in Superconformal
  Chern-Simons Theory and Fundamental Strings in Anti-de Sitter Supergravity
  Dual}}, \href{http://dx.doi.org/10.1088/1126-6708/2009/03/127}{\emph{JHEP}
  {\bfseries 03} (2009) 127},
  [\href{https://arxiv.org/abs/0809.3786}{{\ttfamily 0809.3786}}].

\bibitem{Ouyang:2015iza}
H.~Ouyang, J.-B. Wu and J.-j. Zhang, \emph{{Novel BPS Wilson loops in
  three-dimensional quiver Chern\textendash{}Simons-matter theories}},
  \href{http://dx.doi.org/10.1016/j.physletb.2015.12.021}{\emph{Phys. Lett. B}
  {\bfseries 753} (2016) 215--220},
  [\href{https://arxiv.org/abs/1510.05475}{{\ttfamily 1510.05475}}].

\bibitem{Ouyang:2015bmy}
H.~Ouyang, J.-B. Wu and J.-j. Zhang, \emph{{Construction and classification of
  novel BPS Wilson loops in quiver Chern\textendash{}Simons-matter theories}},
  \href{http://dx.doi.org/10.1016/j.nuclphysb.2016.07.018}{\emph{Nucl. Phys. B}
  {\bfseries 910} (2016) 496--527},
  [\href{https://arxiv.org/abs/1511.02967}{{\ttfamily 1511.02967}}].

\bibitem{Correa:2019rdk}
D.~H. Correa, V.~I. Giraldo-Rivera and G.~A. Silva, \emph{{Supersymmetric mixed
  boundary conditions in AdS$_{2}$ and DCFT$_{1}$ marginal deformations}},
  \href{http://dx.doi.org/10.1007/JHEP03(2020)010}{\emph{JHEP} {\bfseries 03}
  (2020) 010}, [\href{https://arxiv.org/abs/1910.04225}{{\ttfamily
  1910.04225}}].

\bibitem{Ouyang:2022vof}
H.~Ouyang and J.-B. Wu, \emph{{Fermionic Bogomol\`nyi-Prasad-Sommerfield Wilson
  loops in four-dimensional $\mathcal{N}=2$ superconformal gauge theories}},
  \href{http://dx.doi.org/10.21468/SciPostPhys.14.1.008}{\emph{SciPost Phys.}
  {\bfseries 14} (2023) 008},
  [\href{https://arxiv.org/abs/2205.01348}{{\ttfamily 2205.01348}}].

\bibitem{Ouyang:2015ada}
H.~Ouyang, J.-B. Wu and J.-j. Zhang, \emph{{BPS Wilson loops in Minkowski
  spacetime and Euclidean space}},
  \href{http://dx.doi.org/10.1140/epjc/s10052-015-3834-6}{\emph{Eur. Phys. J.
  C} {\bfseries 75} (2015) 606},
  [\href{https://arxiv.org/abs/1504.06929}{{\ttfamily 1504.06929}}].

\bibitem{Siegel:1979wq}
W.~Siegel,
\emph{{Supersymmetric Dimensional Regularization via Dimensional Reduction}},
\href{https://doi.org/10.1016/0370-2693(79)90282-X}{\emph{Phys. Lett. B} {\bfseries 84} (1979), 193-196}

\bibitem{Drukker:2006ga}
N.~Drukker, \emph{{$1/4$ BPS circular loops, unstable world-sheet instantons and
  the matrix model}},
  \href{http://dx.doi.org/10.1088/1126-6708/2006/09/004}{\emph{JHEP} {\bfseries
  09} (2006) 004}, [\href{https://arxiv.org/abs/hep-th/0605151}{{\ttfamily
  hep-th/0605151}}].

\bibitem{Drukker:2006xg}
N.~Drukker and S.~Kawamoto, \emph{{Small deformations of supersymmetric Wilson
  loops and open spin-chains}},
  \href{http://dx.doi.org/10.1088/1126-6708/2006/07/024}{\emph{JHEP} {\bfseries
  07} (2006) 024}, [\href{https://arxiv.org/abs/hep-th/0604124}{{\ttfamily
  hep-th/0604124}}].

\bibitem{Correa:2018fgz}
D.~Correa, M.~Leoni and S.~Luque, \emph{{Spin chain integrability in
  non-supersymmetric Wilson loops}},
  \href{http://dx.doi.org/10.1007/JHEP12(2018)050}{\emph{JHEP} {\bfseries 12}
  (2018) 050}, [\href{https://arxiv.org/abs/1810.04643}{{\ttfamily
  1810.04643}}].

  \bibitem{Correa:2023lsm}
D.~H.~Correa, V.~I.~Giraldo-Rivera and M.~Lagares,
\emph{{Integrable Wilson loops in ABJM: a Y-system computation of the cusp anomalous dimension}},
\href{http://dx.doi.org/10.1007/JHEP06(2023)179}{\emph
{JHEP} \bfseries{06} (2023), 179},
[\href{https://arxiv.org/abs/2304.01924}{{\ttfamily 2304.01924}}].

\bibitem{JKV}
Y.~Jiang, S.~Komatsu and E.~Vescovi, \emph{{to appear}}.

\bibitem{Jiang:2023cdm}
Y.~Jiang, J.-B.~Wu and P.~Yang, \emph{{Wilson-loop One-point Functions in ABJM
  Theory}},  \href{https://arxiv.org/abs/2306.05773}{{\ttfamily 2306.05773}}.

\end{thebibliography}

\providecommand{\href}[2]{#2}\begingroup\raggedright\endgroup

\end{document}